\documentclass[preprint,aps,pre,groupaddress,onecolumn]{revtex4-2} 

\usepackage{natbib}
\usepackage{graphicx}
\usepackage{subfigure}
\usepackage{amssymb}
\usepackage{amsfonts}
\usepackage{amsmath}
\usepackage{amsbsy}
\usepackage{mathrsfs} 
\usepackage{color}
\usepackage{cases}

\usepackage[colorlinks=true,citecolor=blue]{hyperref}
\usepackage{mymacros}

\begin{document} 
\title{Giant superhydrophobic slip of shear-thinning liquids}
\author{Ory Schnitzer}
\author{Prasun K.~Ray}
\affiliation{Department of Mathematics, Imperial College London, London SW7 2AZ, UK}
\begin{abstract}
We theoretically illustrate how complex fluids flowing over superhydrophobic surfaces may exhibit giant flow enhancements in the double limit of small solid fractions ($\epsilon\ll1$) and strong shear thinning ($\beta\ll1$, $\beta$ being the ratio of the viscosity at infinite shear rate to that at zero shear rate). Considering a Carreau liquid within the canonical scenario of longitudinal shear-driven flow over a grooved superhydrophobic surface, we show that, as $\beta$ is decreased, the scaling of the effective slip length at small solid fractions is enhanced from the logarithmic scaling $\ln(1/\epsilon)$ for Newtonian fluids to the algebraic scaling $1/\epsilon^{\frac{1-n}{n}}$, attained for $\beta=\mathcal{O}(\epsilon^{\frac{1-n}{n}})$, $n\in(0,1)$ being the exponent in the Carreau model. We illuminate this scaling enhancement and the geometric-rheological mechanism underlying it through asymptotic arguments and numerical simulations.
\end{abstract}
\maketitle

\newpage


\textit{Introduction}.---Hydrophobic solid surfaces featuring small-scale roughness or an engineered microstructure can spontaneously attain a ``Cassie state'' upon being submerged in liquid, wherein gas pockets are trapped within the surface indentations. Besides exhibiting remarkable wetting properties \cite{Lafuma:03}, such ``superhydrophobic'' surfaces are also promising for drag reduction at small scales 
\cite{Lee:16}. The basis for drag reduction is the mixed conditions experienced by a liquid flowing over a superhydrophobic surface: no-slip on the solid boundaries and, ideally, shear-free on the gas interfaces. As such, a crucial parameter is the solid fraction $\epsilon$ of the compound liquid interface; indeed, in many flow scenarios the hydrodynamic resistance theoretically vanishes as $\epsilon\to 0$ at different rates depending on the surface geometry \cite{Ybert:07}---indicating varying potential for significant drag reductions using small-solid-fraction surfaces. While large flow enhancements have indeed been demonstrated under laboratory conditions \cite{Lee:08}, practical challenges abound including the tendency of the Cassie state to collapse at small solid fractions \cite{Extrand:04} and contamination of the gas interfaces by surfactant molecules, which may result in appreciable deviations from shear-free conditions \cite{Temprano:23}.

Most research on superhydrophobic drag reduction has been focused on Newtonian liquids. Many microscale-flow applications, however, utilize complex fluids \cite{Malkin:18,Rahmani:24}. Shear-thinning liquids are particularly promising in this context. Experimentally, \citet{Gaddam:21} have investigated pressure-driven flow of shear-thinning xanthan gum polymeric solutions through channels with superhydrophobic walls formed of pillars or transverse grooves. At a moderately small solid fraction of $15\%$, they found significant flow enhancements---up to $20$ times larger than those measured for water flowing through the same channels \footnote{Gaddam \textit{et al.}~define the flow enhancement as $Q_t/Q_s-1$, $Q_t$ and $Q_s$ being the flow rates for textured and smooth channels, respectively. For the polymer solutions, they measure enhancements up to $\approx10$ for pillars and up to $\approx 5$ for transverse grooves, in comparison to enhancements $<1$ for water flowing in the same setups.}. Similar flow enhancements were numerically demonstrated by \citet{Gaddam:21} for those configurations, and for longitudinally grooved channels by \citet{Ray:23}. 

Intuitively, giant flow enhancements can be anticipated when \emph{both} the solid fraction is small and the liquid is strongly shear thinning in the sense that the ratio $\beta$ of the viscosity at infinite shear rate to that at zero shear rate is small. 
Indeed, the former (geometric) condition suggests high shear rates surrounding the relatively small solid patches offering viscous resistance; in turn, the latter (rheological)  condition implies low viscosities in those critical regions. Nonetheless, theoretical analyses have so far been confined to studies (of various superhydrophobic setups) in which the rheology is perturbed about the Newtonian case, holding the solid fraction fixed \cite{Crowdy:17ST,Ray:23}. Although these studies have provided valuable analytical insights, in particular highlighting the typical nonmonotonic dependence of the drag reduction upon the externally imposed shear rate \cite{Haase:17,Patlazhan:17,Gaddam:21}, such a perturbative approach is inherently limited to \emph{weak} effects of shear thinning. In practice, many non-Newtonian liquids are strongly shear thinning (e.g., $\beta\approx 0.0007$-$0.01$ for the polymer solutions used by  \citet{Gaddam:21}). 

We here aim to illuminate how strong shear thinning may dramatically affect flows over small-solid-fraction superhydrophobic surfaces. To this end, we  employ scaling arguments and asymptotic  approximations  alongside numerical simulations to study a non-Newtonian version of a prototypical superhydrophobic-flow problem: calculating the effective slip length for shear-driven flow over a longitudinally grooved superhydrophobic surface. We adopt the Carreau model \cite{Bird:book}  in which the fluid viscosity $\eta^*$ monotonically decreases as a function of the shear rate $\dot{\gamma}^*=\sqrt{2\boldsymbol{\mathsf{E}}^*\boldsymbol{:}\boldsymbol{\mathsf{E}}^*}$, where $\boldsymbol{\mathsf{E}^*}$ is the strain-rate tensor, according to the empirical formula 
\begin{equation}\label{eta dim def}
\eta^*=\eta_{\infty}^*+(\eta_0^*-\eta_\infty^*)\left[1+\left(\lambda^*{\dot\gamma}^*\right)^2\right]^{\frac{n-1}{2}}.
\end{equation}
Here $\eta_0^*$ and $\eta_{\infty}^*$ are the limiting viscosities at zero and infinite shear rate, respectively, $\lambda^*$ is a relaxation time whose inverse roughly corresponds to the value of the shear rate about which the viscosity drop occurs, and $n\in(0,1)$ is an exponent controlling the rate of that drop; for intermediately large shear rates, $1\ll \lambda^*\dot{\gamma}^*\ll(\eta_0^*/\eta_{\infty}^*)^{1/(1-n)}$, \eqref{eta dim def} reduces to the model for a power-law fluid in which $\eta^*\propto{\dot{\gamma}^*}{}^{n-1}$. 

\begin{figure}[t!]
\begin{center}
\includegraphics[scale=0.7,trim={0 0.3cm 0 0}]{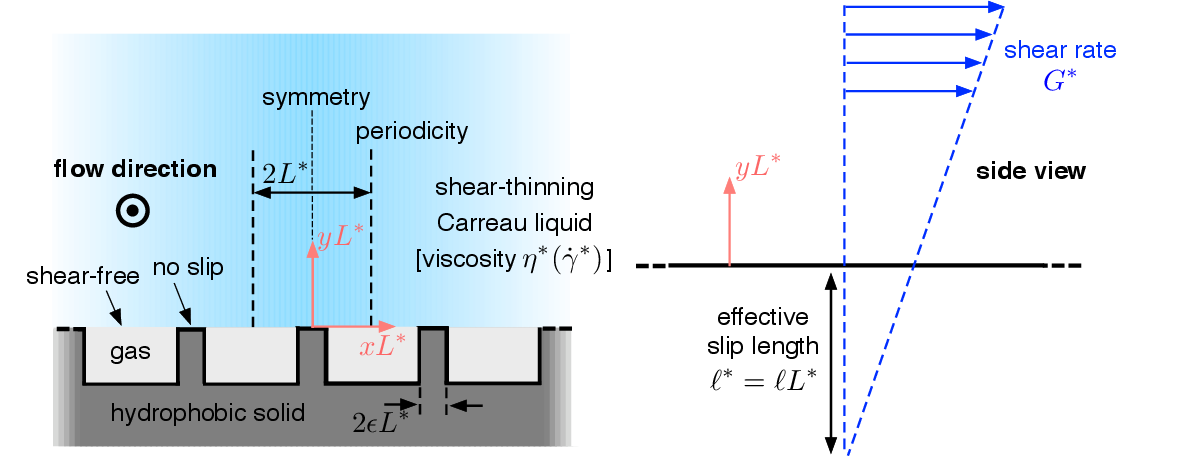}
\caption{Longitudinal shear-driven flow of a Carreau liquid over a grooved superhydrophobic surface.}
\label{fig:schematic}
\end{center}
\end{figure}

\textit{Problem formulation}.---As depicted in Fig.~\ref{fig:schematic}, we consider steady, unidirectional flow of a shear-thinning Carreau liquid [viscosity given by \eqref{eta dim def}] over a grooved solid surface (period $2L_*$, ridge width $2\epsilon L^*$), driven by a simple shear flow (shear rate $G^*$) parallel to the grooves. (An asterisk indicates a dimensional quantity.) We assume a Cassie state where air is trapped within the grooves, and that the menisci are pinned at the corners of the solid ridges, flat and shear-free. The compound liquid interface, henceforth referred to as the superhydrophobic plane, is accordingly flat, consisting of a periodic array of alternating no-slip and shear-free strips representing the solid ridge tops and menisci, respectively. The aspect ratio $\epsilon$ thus corresponds to the solid fraction of the superhydrophobic plane. 

The flow velocity field is parallel to the grooves and invariant along that direction. The sole velocity component therefore satisfies a two-dimensional problem in a plane normal to the grooves. We introduce dimensionless Cartesian coordinates $(x,y)$ and the dimensionless velocity component $w(x,y)$, with lengths normalized by $L^*$ and velocities by $G^*L^*$. The origin is located at the midpoint of an arbitrary ridge top with the $x$ axis lying along the superhydrophobic plane and the $y$ axis pointing into the liquid domain, $y>0$. The periodicity in the $x$ direction, together with the symmetry about $x=0$, allows us to restrict the formulation to the unit-cell $x\in(-1,1)$, with Neumann conditions $\partial w/\partial x=0$ at the cell boundaries $x=\pm1$. 

In the present unidirectional-flow scenario, the Navier--Stokes equations reduce to the momentum equation $\bnabla\bcdot(\eta\bnabla w)=0$, where $\eta=\eta^*/\eta_0^*$ is the ratio of the Carreau viscosity \eqref{eta dim def} to its value at zero shear rate and $\bnabla$ represents the dimensionless gradient operator associated with the Cartesian coordinates $(x,y)$. Furthermore, since $\dot{\gamma}^*=G^*|\bnabla w|$, the dimensionless viscosity $\eta$ can be written as
\begin{equation}\label{eta def}
\eta=\beta+(1-\beta)(1+\mathrm{Cu}^2|\bnabla w|^2)^{\frac{n-1}{2}},
\end{equation} 
where $\beta=\eta_\infty^*/\eta_0^*\in(0,1)$ is the aforementioned ratio of the viscosity at infinite shear rate to that at zero shear rate, and $\mathrm{Cu}=\lambda^*G^*\ge0$ is the Carreau number characterizing the imposed shear rate.  At $y=0$, the velocity satisfies mixed-boundary conditions: no-slip, $w=0$, on the ridge top, $|x|<\epsilon$; and shear-free, $\partial w/\partial y=0$, on the adjacent menisci, $\epsilon<|x|<1$. Lastly, we have the far-field condition $w\sim y$ as $y\to\infty$, representing the approach to the imposed linear flow. 

The far-field expansion of the velocity can be extended as  
\begin{equation}\label{far field extended}
w= y+\ell + o(1) \quad \text{as} \quad y\to\infty,
\end{equation}
in which $\ell$ is the effective slip length (normalized by $L^*$). According to \eqref{far field extended}, the superhydrophobic patterning induces a uniform stream of dimensional magnitude $G^*L^*\ell$ at large distances from the surface. Thus, $\ell$ measures the ``flow enhancement'' associated with superhydrophobicity; it may also be interpreted as the dimensionless distance below the superhydrophobic plane where the far-field velocity, namely the imposed shear flow plus the induced uniform stream, vanishes (Fig.~\ref{fig:schematic}b). Accordingly, our focus will be to identify and characterize parameter regimes in which $\ell$ is amplified. 

From the above problem formulation, it is straightforward to derive the integral relation 
\begin{equation}\label{integral}
\int_{-\epsilon}^{\epsilon}\left(\eta\pd{w}{y}\right)_{y=0}\mathrm{d}x=2\bar{\eta},
\end{equation}
wherein 
\begin{equation}\label{bar eta}
\bar{\eta}=\beta+(1-\beta)(1+\mathrm{Cu}^2)^{\frac{n-1}{2}}
\end{equation}
is the far-field limit of the dimensionless viscosity \eqref{eta def}. Physically, \eqref{integral} represents a balance between the viscous stresses acting on the solid ridge tops and the shear stresses imposed at large distances from the surface. While \eqref{integral} is not independent of the problem formulation it will prove invaluable to our analysis.

\textit{Small solid fractions}.---In the Newtonian case, wherein $\eta$ is constant, \citet{Philip:72} solved the above boundary-value problem for $w$ using complex-variable methods. The small-solid-fraction expansion of his closed-form expression for the effective slip length, 
\begin{equation}\label{small eps newton}
\ell = \frac{2}{\pi}\ln \sec\frac{\pi(1-\epsilon)}{2}=\frac{2}{\pi}\ln \frac{1}{\epsilon}+ \frac{2}{\pi}\ln \frac{2}{\pi} + o(1) \quad \text{as} \quad \epsilon\to0,
\end{equation}
reveals a singular logarithmic scaling in that limit. 
How is this behavior modified by shear thinning? 

\citet{Schnitzer:17} showed that the logarithmic approximation in \eqref{small eps newton} can be derived without the need for detailed calculation. His argument can be generalized as follows to the present case of a shear-thinning liquid. For $\epsilon\ll1$ (holding the rheological parameters fixed), we expect $|\bnabla w|\gg1$ in the vicinity of the ridge top. The fluid is thus approximately Newtonian in that region with $\eta\sim \beta$ [cf.~\eqref{eta def}], whereby the force balance  \eqref{integral} can be approximated as  
\begin{equation}\label{integral small eps}
\beta \int_{-\epsilon}^{\epsilon}\left(\pd{w}{y}\right)_{y=0}\mathrm{d}x \approx  2\bar{\eta}.
\end{equation} 
At intermediate radial distances, $\epsilon\ll r\ll1$ ($r$ being a radial coordinate measured from the origin)---large enough such that the ridge top effectively shrinks to the origin, yet still small so that the fluid remains Newtonian and the unit-cell boundaries are effectively at infinity---\eqref{integral small eps} implies the velocity approximation  
\begin{equation}\label{w intermediate}
w =  \frac{2\bar{\eta}}{\pi\beta}\ln \frac{r}{\epsilon} + o(\ln  r), 
\end{equation}
corresponding to the flow due to a line force at the origin directed parallel to itself \footnote{Following \citet{Schnitzer:17}, it is convenient to consider a mathematical analogy wherein $\bnabla w$ is interpreted as a fictitious ideal flow, which emanates from the ridge tops at a rate determined by \eqref{integral small eps}. In that analogy, \eqref{w intermediate} represents a two-dimensional potential source flow.}.
To leading order, the intermediate behavior \eqref{w intermediate} gives the logarithmically large uniform-velocity approximation $w\sim (2\bar{\eta}/\pi\beta)\ln (1/\epsilon)$. In fact, this uniform-velocity approximation extends to $r=\text{ord}(1)$, since (i) the far-field shear flow is only of order unity and (ii) $w$ satisfies homogeneous Neumann conditions on the menisci and unit-cell boundary. 
The far-field expansion \eqref{far field extended} accordingly yields the slip-length approximation 
\begin{equation}\label{log approximation}
\ell \sim \frac{2}{\pi}\left[1+\frac{1-\beta}{\beta}(1+\mathrm{Cu}^2)^{\frac{n-1}{2}}\right]\ln\frac{1}{\epsilon} \quad \text{as} \quad \epsilon\to 0,
\end{equation}
where we have substituted \eqref{bar eta} for $\bar{\eta}$. 

As in the Newtonian case [cf.~\eqref{small eps newton}], the leading-order logarithmic approximation \eqref{log approximation} is subject to an $\text{ord}(1)$ correction term---formally negligible, yet numerically significant for realistically small $\epsilon$. In the supplementary material (SM) \footnote{\label{sm ref}The supplementary material can be obtained from the authors (o.schnitzer@imperial.ac.uk).}, we use matched asymptotics \cite{Hinch:91} to derive and analyze the flow problem governing that correction term.
 
 The logarithmic approximation \eqref{log approximation} generally exceeds its Newtonian analog, $(2/\pi)\ln(1/\epsilon)$ [cf.~\eqref{small eps newton}]. Naively, one might expect the former to approach the latter in any limit of the rheological parameters corresponding to a Newtonian liquid (of arbitrary viscosity, given that  the Newtonian slip length is independent of viscosity). This is evidently the case for $\beta\to1$, or for $\mathrm{Cu}\to\infty$, but not for $\mathrm{Cu}\to0$ or $n\to1$! In particular, the approximation \eqref{log approximation} decreases monotonically with $\mathrm{Cu}$ from its maximum at $\mathrm{Cu}=0$, which is  $1/\beta$-times larger than the Newtonian approximation. The resolution to this apparent paradox is that \eqref{log approximation} fails in the weak-shear regime $\mathrm{Cu}=\mathcal{O}(\epsilon)$, where the viscosity is no longer Newtonian in the vicinity of the ridge tops. Of course, $\ell$ must drop to its Newtonian value for sufficiently small $\mathrm{Cu}$, implying that $\ell$ exhibits a maximum, say $\ell_{max}$, as a function of $\mathrm{Cu}$. Subtly, an analysis in the intermediate limit $\epsilon\ll\mathrm{Cu}\ll1$ (see SM) shows that the location of the maximum $\mathrm{Cu}_{max}=\text{ord}(1/\sqrt{\ln(1/\epsilon)})$, whereby the value of \eqref{log approximation} at $\mathrm{Cu}=0$ approximates $\ell_{max}$. (The limit $n\to1$ involves a similar subtlety.)

\textit{Small solid fractions and strong shear thinning.}---The leading-order approximation \eqref{log approximation} exhibits a weak (logarithmic) divergence of the slip length in the small-solid-fraction limit, comparable to that in the Newtonian case. In the double limit of small solid fractions and strong shear thinning, however, it predicts a more dramatic (algebraic) enhancement: 
\begin{equation}\label{double limit}
\ell \sim \frac{2}{\pi}(1+\mathrm{Cu}^2)^{\frac{n-1}{2}}\times \frac{1}{\beta} \ln \frac{1}{\epsilon} \quad \text{as} \quad \epsilon\to0, \quad  \beta \to0.
\end{equation}
We stress that this dramatic enhancement requires that \emph{both} $\epsilon$ and $\beta$ are small: the small solid fraction results in an increased shear rate in the vicinity of the ridge top, whereby strong shear thinning implies a small viscosity in that region; the force balance \eqref{integral} then dictates an amplified flow. Conversely, if $\beta\to0$ with $\epsilon$ held fixed, the viscosity becomes asymptotically small only local to the ridge corners; the force integral is thus dominated by the bulk of the ridge top where the fluid retains moderate viscosity. We therefore anticipate that $\lim_{\beta\to0}\ell$ exists for any fixed $\epsilon>0$; i.e., strong shear thinning by itself does not lead to any singular slip enhancement. 

While approximation \eqref{double limit} gives some indication of the remarkable slip enhancement that can be achieved when both $\epsilon$ and $\beta$ are small, it is important to recognize that it has been derived by considering an ordered limit process: $\epsilon\to0$ followed by $\beta\to0$. We can see that the order of limits matters, as our preceding arguments fail for sufficiently small $\beta$. Indeed, the small-solid-fraction approximation \eqref{log approximation} assumes that the shear rate in the $\mathcal{O}(\epsilon)$ vicinity of the ridge top is so large that the liquid attains its Newtonian infinite-shear limit there. On the one hand, \eqref{integral small eps} suggests that the scaling of the shear rate $|\bnabla w|$ is enhanced from $1/\epsilon$ to $1/(\epsilon\beta)$. On the other hand, neglecting the second term in the viscosity model \eqref{eta def} now requires $\beta|\bnabla w|^{1-n}\gg 1$, showing that the assumption $\eta\sim\beta$ fails for $\beta =\mathcal{O}(\epsilon^{\frac{1-n}{n}})$. Under that distinguished scaling, \eqref{double limit} suggests that $\ell$ adopts the algebraically large scaling $\epsilon^{\frac{n-1}{n}}$.

\textit{Distinguished strong-shear-thinning limit}.---We conclude that, as $\beta$ is decreased, the small-solid-fraction singularity transitions from a logarithmic regime, where \eqref{log approximation} holds, to an algebraic one, where we posit the approximation 
\begin{equation}
\ell \sim \epsilon^{\frac{n-1}{n}}\tilde{\ell}(\tilde{\beta},\mathrm{Cu},n) \quad \text{as} \quad \epsilon\to0, \quad \text{with} \quad \tilde{\beta}=\epsilon^{\frac{n-1}{n}}\beta \quad \text{fixed}.
\label{slip expansion distinguished}
\end{equation}
We next formulate a problem governing the rescaled leading-order slip length $\tilde{\ell}$, which we shall see depends upon $\epsilon$ solely via the rescaled viscosity ratio $\tilde{\beta}$. To this end, we shall  asymptotically match an ``outer'' approximation corresponding to the order-unity scale of the unit-cell and an ``inner'' approximation corresponding to the $\mathcal{O}(\epsilon)$ vicinity of the ridge top. In fact, the leading-order outer approximation is simply the uniform flow $w\sim \epsilon^{\frac{n-1}{n}}\tilde{\ell}$. 
\begin{figure}[t!]
\begin{center}
\includegraphics[scale=0.48,trim={3.5cm 1cm 3cm 0}]{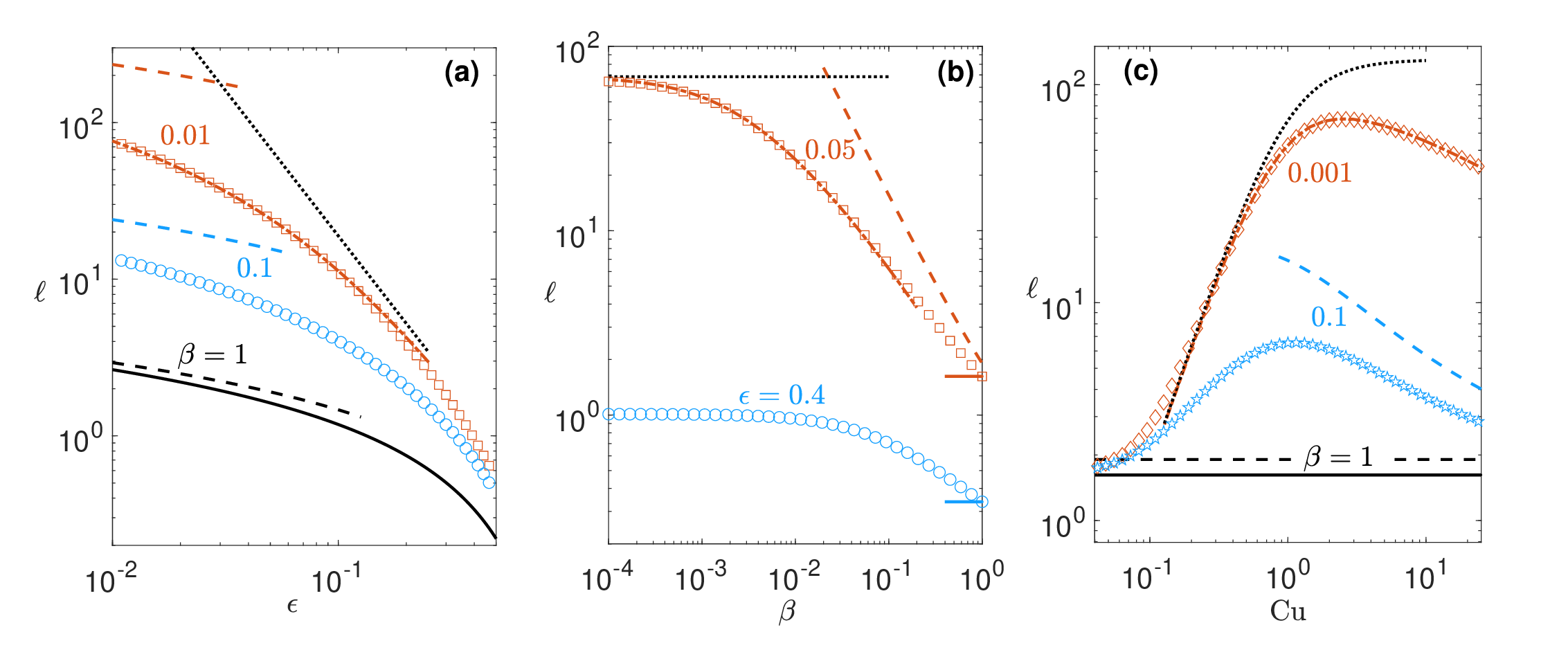}
\caption{Dimensionless slip length $\ell$ as a function of (a) solid fraction $\epsilon$, for $\mathrm{Cu}=1$ and the indicated values of the viscosity ratio $\beta$; (b) $\beta$, for $\mathrm{Cu}=1$ and the indicated values of $\epsilon$; (c) Carreau number $\mathrm{Cu}$, for $\epsilon=0.05$ and the indicated values of $\beta$. In all cases, $n=0.35$. Symbols: numerical solutions of the exact problem. Dashed curves: logarithmic approximation \eqref{log approximation} as $\epsilon\to0$. Dash-dotted curves: Numerical solution to the inner problem in the distinguished limit $\beta=\mathcal{O}(\epsilon^\frac{1-n}{n})$ [cf.~\eqref{slip expansion distinguished}]. Dotted curves: small-$\epsilon$ approximation \eqref{limiting slip} for $\lim_{\beta\to0}\ell$, with $\hat{\ell}(0.35)\approx 0.50$. The solid curves depict the exact Newtonian solution [cf.~\eqref{small eps newton}].}
\label{fig:numericsA}
\end{center}
\end{figure}

To formulate the inner problem, we introduce the stretched position vector $\tilde{\bx}=\bx/\epsilon$, wherein $\tilde{\bx}=(\tilde{x},\tilde{y})$ and $\bx=(x,y)$, and corresponding radial coordinate $\tilde{r}=r/\epsilon$. Under this rescaling, the liquid domain becomes the half-plane $\tilde{y}>0$, with the liquid interface $\tilde{y}=0$ composed of the ridge top ($|\tilde{x}|<1$) and neighbouring menisci ($|\tilde{x}|>1$). The inner velocity field is expanded as $w\sim \epsilon^{\frac{n-1}{n}}\tilde{w}(\tilde{\bx})$. The rescaled velocity $\tilde{w}$ satisfies the momentum equation $\tilde{\bnabla}\bcdot\left(\tilde{\eta}\tilde{\bnabla}\tilde{w}\right)=0$, 
wherein $\tilde{\bnabla}$ is the gradient operator with respect to $\tilde{\bx}$ and 
\begin{equation}
\tilde{\eta}=\tilde{\beta}+\mathrm{Cu}^{n-1}|\tilde{\bnabla}\tilde{w}|^{n-1}
\label{tilde viscosity}
\end{equation}
is the leading-order approximation for the viscosity $\eta$ (rescaled by $\epsilon^{\frac{1-n}{n}}$). At $\tilde{y}=0$, the velocity field satisfies mixed-boundary conditions: no-slip, $\tilde{w}=0$, for $|\tilde{x}|<1$; and shear-free, $\partial\tilde{w}/\partial\tilde{y}=0$, for $|\tilde{x}|>1$. 
Matching with the outer uniform flow suggests the far-field condition
\begin{equation}
\tilde{w} \to \tilde{\ell} \quad \text{as} \quad \tilde{r} \to \infty.
\label{tilde far}
\end{equation}
Lastly, the inner problem is closed by the integral constraint
\begin{equation}
\int_{-1}^1 \left(\tilde{\eta}\pd{\tilde{w}}{\tilde{y}}\right)_{\tilde{y}=0}\mathrm{d}\tilde{x}= 
2(1+\mathrm{Cu}^2)^{\frac{n-1}{2}},
\label{tilde integral}
\end{equation}
which follows from a leading-order balance of \eqref{integral}. In contrast to the exact formulation, wherein \eqref{integral} was a derived relation, here \eqref{tilde integral} provides independent information serving to determine $\tilde{\ell}$ (circumventing asymptotic matching at higher orders).  

While in general the inner problem needs to be solved numerically, scaling results and leading-order approximations can readily be derived in limiting cases. We first assume $\tilde{\beta}\ll\mathrm{Cu}^{n-1}|\tilde{\bnabla}\tilde{w}|^{n-1}$, such that the viscosity model \eqref{tilde viscosity} reduces to that for a power-law fluid. We may then factor out $\mathrm{Cu}$ by writing $(\tilde{w},\tilde{\ell})=\chi(\hat{w},\hat{\ell})$, wherein $\chi=[\mathrm{Cu}^2/(1+\mathrm{Cu}^2)]^{(1-n)/2n}$, such that $\hat{w}=\hat{w}(\tilde{\bx};n)$ and $\hat{\ell}=\hat{\ell}(n)$. Using the estimate $|\tilde{\bnabla}\tilde{w}|\approx \chi$, we see that the assumed smallness of $\tilde{\beta}$ is consistent for $\tilde{\beta}\mathrm{Cu}^{(1-n)/n}\ll1$ with $\mathrm{Cu}\ll1$, or for $\tilde{\beta}\mathrm{Cu}^{1-n}\ll1$ with $1/\mathrm{Cu}=\mathcal{O}(1)$. We accordingly find the approximations 
\begin{subnumcases} {\label{inner approximation 1} \tilde{\ell} \sim \hat{\ell}(n) \times  }
\left(\frac{\mathrm{Cu}^2}{1+\mathrm{Cu}^2}\right)^{\frac{1-n}{2n}}, & for $\tilde{\beta}\mathrm{Cu}^{1-n}\ll1, \,\,\, 1/\mathrm{Cu}=\mathcal{O}(1)$, \label{inner approximation 1 a}\\
\mathrm{Cu}^{\frac{1-n}{n}}, & for $\tilde{\beta}\mathrm{Cu}^{\frac{1-n}{n}}\ll1, \,\,\, \mathrm{Cu}\ll1$, \label{inner approximation 1 b}
\end{subnumcases}
wherein $\hat{\ell}(n)$ is governed by the problem for $\hat{w}$, whose formulation can be found in the SM.  

\begin{figure}[t!]
\begin{center}
\includegraphics[scale=0.5,trim={3.3cm 1cm 3cm 0}]{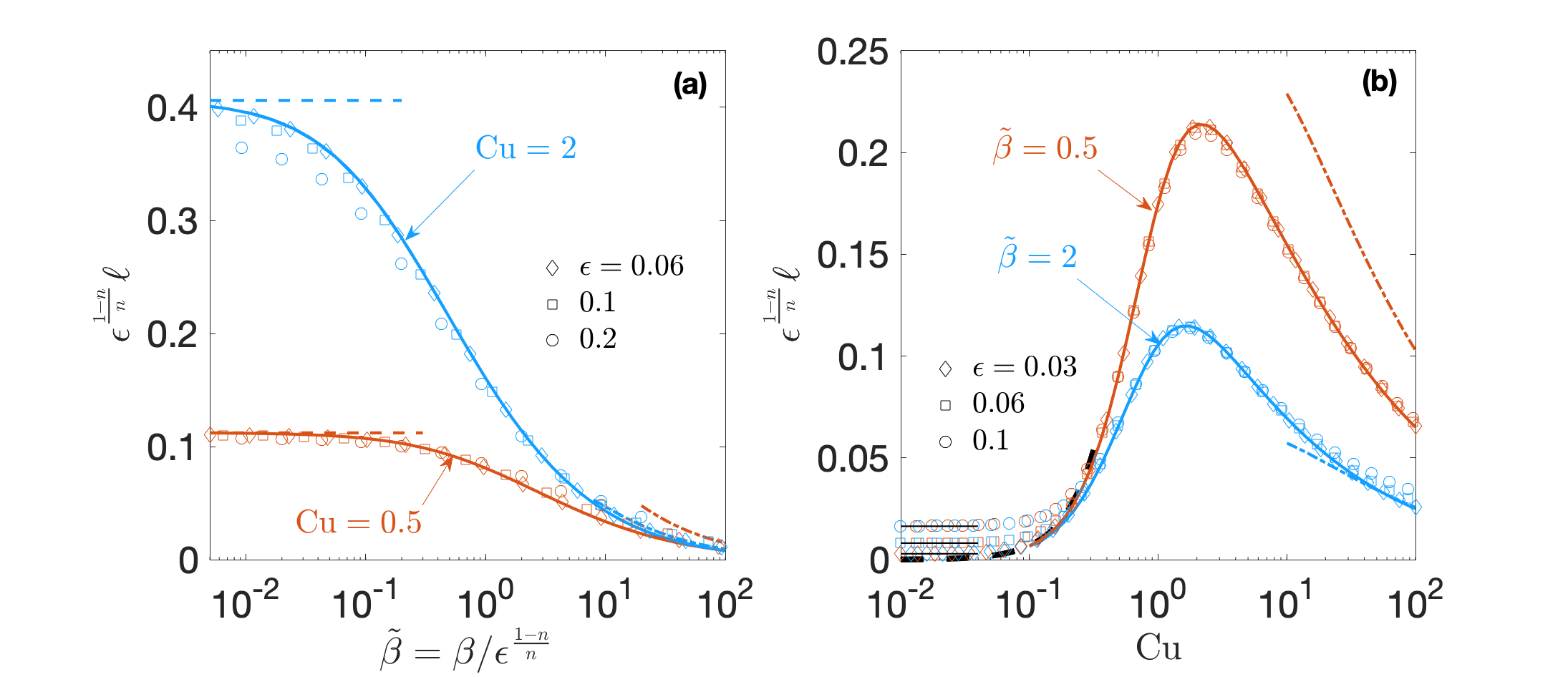}
\caption{Rescaled slip length $\epsilon^{\frac{1-n}{n}}\ell$ as a function of (a) rescaled viscosity ratio $\tilde{\beta}=\beta/\epsilon^{\frac{1-n}{n}}$, for the indicated values of the Carreau number $\mathrm{Cu}$; (b) $\mathrm{Cu}$, for the indicated values of $\tilde{\beta}$. In all cases, $n=0.35$. Symbols: numerical solutions of exact problem, for the indicated values of $\epsilon$. Solid curves: $\tilde{\ell}(\tilde{\beta},\mathrm{Cu},n)$, obtained by numerical solution to the inner problem in the distinguished limit \eqref{slip expansion distinguished}. In (a), the dashed curves depict the small-$\tilde{\beta}$ approximation $\hat{\ell}(n)[\mathrm{Cu}^2/(1+\mathrm{Cu}^{2})]^{(1-n)/(2n)}$, wherein $\hat{\ell}(0.35)\approx0.50$ [cf.~\eqref{inner approximation 1 a}]; and the dash-dotted curves the large-$\tilde{\beta}$ approximation $2n/[\pi(1-n)](1+\mathrm{Cu}^2)^{(n-1)/2}\tilde{\beta}^{-1}\ln\tilde{\beta}$  [cf.~\eqref{inner approximation 2 a}]. In (b), the dashed curve depicts the small $\mathrm{Cu}$ approximation $\hat{\ell}(n)\mathrm{Cu}^{(1-n)/n}$ [cf.~\eqref{inner approximation 1 b}]; the dash-dotted curves the large-$\mathrm{Cu}$ approximation $(2n/\pi)(1+\mathrm{Cu}^2)^{(n-1)/2}\tilde{\beta}^{-1}\ln\mathrm{Cu}$ [cf.~\eqref{inner approximation 2 b}]; and the solid curves the exact Newtonian solution [cf.~\eqref{small eps newton}].}
\label{fig:numericsB}
\end{center}
\end{figure}

We next trial the assumption $\tilde{\beta}\gg\mathrm{Cu}^{n-1}|\tilde{\bnabla}\tilde{w}|^{n-1}$, such that the viscosity model \eqref{tilde viscosity} reduces to the Newtonian approximation $\tilde{\eta}\sim\tilde{\beta}$. The integral condition \eqref{tilde integral} then suggests that $|\tilde{\bnabla}\tilde{w}|$ scales as $\tilde{\beta}^{-1}(1+\mathrm{Cu}^2)^{(n-1)/2}$, which appears consistent for $\tilde{\beta}\mathrm{Cu}^{(1-n)/n}\gg1$ with $\mathrm{Cu}=\mathcal{O}(1)$, or for $\tilde{\beta}\mathrm{Cu}^{1-n}\gg1$ with $\mathrm{Cu}\gg1$. However, the integral condition \eqref{tilde integral} in conjunction with the Newtonian behavior imply $\tilde{w}\sim (2/\pi\tilde{\beta})(1+\mathrm{Cu}^2)^{(n-1)/2}\ln \tilde{r}$ as $\tilde{r}\to\infty$, which is incompatible with the far-field condition  \eqref{tilde far}. The resolution to this inconsistency is that non-Newtonian effects do become important at distances $\tilde{r}$ of order $\tilde{\beta}^{n/(1-n)}(1+\mathrm{Cu}^2)^{(n-1)/2}\mathrm{Cu}$, and can be shown to cut off that logarithmic growth. This implies the (only logarithmically accurate) approximations \footnote{A higher order approximation as $\tilde{\beta}\to\infty$ is derived in the SM.} 
\begin{subnumcases} {\label{inner approximation 2} \tilde{\ell} \sim \frac{2}{\pi}(1+\mathrm{Cu}^2)^{\frac{n-1}{2}}\tilde{\beta}^{-1}\ln  }
\tilde{\beta}^{\frac{n}{1-n}}\mathrm{Cu}, & for $\tilde{\beta}\mathrm{Cu}^{\frac{1-n}{n}}\gg1, \,\,\, \mathrm{Cu}=\mathcal{O}(1)$, \label{inner approximation 2 a}\\
\tilde{\beta}^{\frac{n}{1-n}}\mathrm{Cu}^n, & for $\tilde{\beta}\mathrm{Cu}^{1-n}\gg1, \,\,\, \mathrm{Cu}\gg1$ \label{inner approximation 2 b}.
\end{subnumcases}

Approximation \eqref{inner approximation 1 a} implies the result 
\begin{equation}\label{limiting slip}
\lim_{\beta\to0}\ell \sim \frac{\hat{\ell}(n)}{\epsilon^{\frac{1-n}{n}}}\left(\frac{\mathrm{Cu}^2}{1+\mathrm{Cu}^2}\right)^{\frac{1-n}{2n}} \quad \text{as} \quad \epsilon\to0,
\end{equation} 
which (assuming $\ell$ is monotonic in $\beta$) gives a small-solid-fraction approximation for the maximal attainable slip length possible by means of reducing $\beta$. Furthermore, approximation \eqref{inner approximation 1 a} shows that $\tilde{\ell}$ vanishes as $\mathrm{Cu}\to0$, scaling as $\mathrm{Cu}^{(1-n)/n}$ in that limit. On the other hand, approximation \eqref{inner approximation 2 b} shows that $\tilde{\ell}$ also vanishes as $\mathrm{Cu}\to\infty$, scaling as $\mathrm{Cu}^{n-1}\ln\mathrm{Cu}$ in that limit. Accordingly, the scaling of  $\mathrm{Cu}_{\text{max}}$ increases to $\text{ord}(1)$ in the distinguished limit \eqref{slip expansion distinguished}, with $\ell_{max}=\text{ord}(\epsilon^{(n-1)/n})$. Lastly, approximation \eqref{inner approximation 2 a} confirms that $\tilde{\ell}$ vanishes as $\tilde{\beta}\to\infty$, scaling as $\tilde{\beta}^{-1}\ln \tilde{\beta}$ and matching with \eqref{double limit} in that limit. 

\textit{Numerical results}.---To illustrate the above findings, we have numerically solved the exact formulation governing the velocity field $w(\bx)$ and slip length $\ell$, as well as the inner problem governing the rescaled velocity field $\tilde{w}(\tilde{\bx})$ and slip length $\tilde{\ell}$ in the distinguished limit \eqref{slip expansion distinguished}. The exact problem was solved using a spectral scheme similar to that employed by \citet{Ray:23} in the case of pressure-driven channel flow. In that scheme, the local singular behaviour of the velocity field near the ridge-top edges is analytically subtracted. Since the viscosity approaches its infinite-shear limit as the edges are approached, the local behavior is the same as for a Newtonian fluid---the velocity on the menisci vanishing with the square-root of the distance from the edges. To solve the inner problem, we map the domain to a strip using elliptical coordinates, which effectively removes the singularities at the ridge-top edges and provides an exponential stretching in the radial direction (see SM). The mapped problem is then solved in Matlab using the Partial Differential Equations Toolbox \cite{PDET}, in conjunction with the \texttt{fsolve} routine applied to the integral constraint \eqref{tilde integral}. 

Fig.~\ref{fig:numericsA} presents the variation of the dimensionless slip length $\ell$ as a function of the solid fraction $\epsilon$ and the rheological parameters $\beta$ and $\mathrm{Cu}$, for $n=0.35$. Numerical solutions of the exact problem are compared with the logarithmic (i.e., grossly inaccurate) approximation \eqref{log approximation} for $\epsilon\ll1$; the solution to the inner problem in the distinguished limit \eqref{slip expansion distinguished}, which is seen to provide highly accurate approximations when both $\epsilon$ and $\beta$ are small; the exact Newtonian solution [cf.~\eqref{small eps newton}]; and the approximation \eqref{limiting slip} for $\lim_{\beta\to0}\ell$. In the latter approximation, we obtain $\hat{\ell}(0.35)\approx 0.50$ by substituting a numerical estimate of $\lim\tilde{\ell}_{\tilde{\beta}\to0}$ into \eqref{inner approximation 1} \footnote{Rather than directly solving the reduced problem for $(\hat{w},\hat{\ell})$, which, as shown in the SM, entails a modified local behavior approaching the ridge-top edges.}. As predicted by the theory, the rate at which $\ell$ grows as $\epsilon\to0$ (without bound) increases as $\beta$ is decreased; for a small value of $\beta$, the growth as $\epsilon\to0$ passes through an algebraic phase [in accordance with the distinguished approximation \eqref{slip expansion distinguished}] before recovering the logarithmic dependence predicted in the limit $\epsilon\to0$. Furthermore, we observe that $\ell$ plateaus as $\beta\to0$, with $\epsilon$ fixed, and that $\ell$ exhibits a maximum as a function of $\mathrm{Cu}$. In Fig.~\ref{fig:numericsB}, we demonstrate that the solution of the inner problem in the distinguished limit \eqref{slip expansion distinguished} allows to collapse numerical values for small $\epsilon$ and $\beta$ on $\epsilon$-independent curves. 

\textit{Concluding remarks}.---We have used asymptotic arguments and numerical simulations to theoretically illuminate the mechanism by which the \emph{combination} of small solid fractions and strong shear thinning can enable giant effective slip lengths for flows over superhydrophobic surfaces; the practical importance of this mechanism is that it enables large flow enhancements (or drag reductions) at rather moderately small solid fractions, as indeed has been demonstrated experimentally by \citet{Gaddam:21} for pressure-driven flow. While we have focused on the canonical problem of a grooved superhydrophobic surface in longitudinal shear flow, it is clear that this geometric-rheological mechanism and the sort of asymptotic arguments we have employed to describe it may be generalized to describe many other superhydrophobic flow configurations. 


\bibliography{refs}
\end{document}